\documentclass[a4paper,USenglish,cleveref,thm-restate]{lipics-v2021}
\usepackage{xspace}
\usepackage{xcolor}
\usepackage{booktabs}
\usepackage{nicefrac}
\usepackage[disableredefinitions]{complexity}
\usepackage{thmtools}

\renewcommand{\leq}{\leqslant}
\renewcommand{\geq}{\geqslant}

\hideLIPIcs
\nolinenumbers

\patchcmd{\acknowledgementsHeading}{Acknowledgements}{Acknowledgments}{}{}

\title{A Simple Computer-Assisted Proof of the 5/6-Bound for Real-Period Pinwheel Scheduling}
\titlerunning{A Computer-Assisted Proof of the 5/6-Bound}
\author{Hiroshi Fujiwara}{Shinshu University, Nagano, Japan \and \url{http://www.cs.shinshu-u.ac.jp/~fujiwara/}}{fujiwara@shinshu-u.ac.jp}{https://orcid.org/0009-0002-9711-3660}{}
\author{Grzegorz Gutowski}{Institute of Theoretical Computer Science, Faculty of Mathematics and Computer Science, Jagiellonian University, Krak\'ow, Poland \and \url{https://grzegorz.gutowscy.pl}}{grzegorz.gutowski@uj.edu.pl}{https://orcid.org/0000-0003-3313-1237}{Partially supported by grant no.~2023/49/B/ST6/01738 from the National Science Centre, Poland.}
\author{Katsuhisa Ouchi}{Shinshu University, Nagano, Japan \and \url{https://soar-rd.shinshu-u.ac.jp/search/detail.html?lang=ja&systemId=jVTpbUkh}}{k_ouchi@shinshu-u.ac.jp}{}{}
\author{Alexander Wolff}{Universität Würzburg, Germany \and \url{https://www.informatik.uni-wuerzburg.de/en/algo/team/wolff-alexander/}}{}{https://orcid.org/0000-0001-5872-718X}{}
\authorrunning{H. Fujiwara, G. Gutowski, K. Ouchi, and A. Wolff}

\newcommand{\N}{\ensuremath{\mathbb{N}}\xspace}
\newcommand{\dens}{\mathrm{D}}
\newcommand{\pot}{\mathrm{P}}
\newcommand{\densth}{\ensuremath{\delta}}

\newcommand{\densities}{\mathbf{d}}

\newcommand{\code}[1]{\texttt{#1}}

\graphicspath{{./figures}}

\hypersetup{hidelinks,bookmarksopen=true,bookmarksopenlevel=2,bookmarksnumbered=true}
\begin{document}

\maketitle

\begin{abstract}
  In pinwheel scheduling with real periods, an instance consists of $n$ tasks with real periods $1\leq a_1\leq a_2\leq\cdots\leq a_n$.
  The goal is to find a schedule that assigns one task to each day indefinitely such that, for every task~$i \in \{1,\dots,n\}$ and every positive integer~$\ell$, task~$i$ occurs at least $\ell$ times in every block of $\lceil\ell a_i\rceil$ consecutive days.
  Given an instance $(a_1,a_2,\ldots,a_n)$,
  its \emph{dimension} is~$n$,
  its \emph{density} is $\sum_{i=1}^n {1}/{a_i}$, and
  its \emph{potential} is $\sum_{i=1}^{n-1} {1}/{a_i}$.

  We prove that every real-valued instance of density at most $\nicefrac{5}{6}$ admits a periodic schedule of period length less than $4\cdot2^n$.
  For integer periods, the $\nicefrac{5}{6}$ density bound was a long-standing conjecture settled by Kawamura using a computer-assisted proof~[STOC 2024, PNAS 2026].
  Our result confirms the real-period conjecture of Fujiwara, Miyagi, and Ouchi~[DMTCS 2026].

  Our computer-assisted proof builds on Kawamura's ideas while using fewer explicit schedules.
  It combines dimension reduction by folding with nine satisfiability modulo theories (SMT) checks over linear real arithmetic.
  A set of 14\,287 explicit periodic schedules contains a schedule for every instance of dimension at most~$8$ and density at most $\nicefrac{5}{6}$, and for every instance of dimension~$9$ and potential at most $\nicefrac{5}{6}$.
  Folding is used to reduce every instance of larger dimension and potential at most $\nicefrac{5}{6}$ to an instance of dimension~$9$ and the same potential bound.

  We also prove that every real-valued instance with minimum period at least~$4$ and density at most $\nicefrac{17}{20}=0.85$ admits a periodic schedule of period length less than $19\cdot2^n$.
  The density bound $\nicefrac{17}{20}$ is tight and improves the $0.84$-bound of Mishra, Rho, and Kleinberg~[arXiv 2025] for integer periods.
  It also resolves the $a_1=4$ case of Kawamura's 2025 conjecture that integer-valued instances of density at most $1-\frac{a_1-1}{a_1(a_1+1)}$ are schedulable.
\supplementdetails[linktext=Source code and certificates]{Software}{https://github.com/grzegorz-gutowski/pinwheel_scheduling}
\acknowledgements{We thank Nuntiatio Collective and Fabian Fetik for communicating their work on the real-period conjecture.
Their use of the folding potential motivated the reduction of our schedule-domination computations to a domain with a bounded potential.
We are grateful to Akitoshi Kawamura for useful discussions.
This work has benefited from \href{https://www.dagstuhl.de/26412}{Dagstuhl Seminar~26412}, ``Recent Trends in Perpetual Scheduling'' (October 4--9, 2026).}
\end{abstract}

\section{Introduction}
\label{sec:introduction}

In the classical pinwheel scheduling problem, we are given $n$ tasks and a
positive integer period $a_{i}$ for each task $i \in [n] = \{1, 2,
\ldots, n\}$.  The goal is to find a schedule that performs one task
per day such that each task $i$ is performed at least once every
$a_{i}$ consecutive days.
Let~\N be the set of all positive integers.
Formally, an integer-valued \emph{instance} of the
pinwheel scheduling problem is a non-decreasing sequence $A = (a_1,a_2,\ldots,a_n) \in
\N^{n}$, where $n$ is the \emph{dimension} of the instance and
each integer $a_{i}$ is the \emph{period} of task~$i$. 
A \emph{schedule} is a map $S \colon \N \to [n]$, where $S(t)$ is the task performed on day~$t$.
We seek a schedule satisfying the following condition for each task $i \in [n]$:
\begin{quote}
  for each $m \in \N$, there exists a day $t \in [m,
    m + a_{i}) \cap \N$ such that $S(t) = i$.
\end{quote}
A schedule is said to be \emph{valid} if it satisfies the condition
above, and an instance is said to be \emph{schedulable} if there
exists a valid schedule for the instance.

For an instance $A = (a_1,a_2,\ldots,a_n)$, 
the \emph{density} of task~$i$ is $d_i=\nicefrac{1}{a_i}$, and $\densities = (d_1,d_2,\dots,d_n)$ is the \emph{density vector} of $A$.
The \emph{density} of $A$ is defined as $\dens(A) = \sum_{i=1}^{n} d_i$, which intuitively
indicates the minimum average number of task executions required per day.
Since only one task is performed per day, every schedulable instance $A$ satisfies $\dens(A)\leq1$.
The \emph{potential} of $A$ is defined as $\pot(A) = \sum_{i=1}^{n-1} d_i$.
We have $\pot(A) = \dens(A) - d_n$, where $d_n$ is the smallest task density.
In particular, $\pot(A) \leq \dens(A)$ holds for any instance $A$.
The potential will play a crucial role in our analysis.
The \emph{density threshold} is the largest real number~$\densth$ such that every
instance whose density does not exceed~$\densth$ is schedulable.

For a simple example, observe that the instance $(2,3,a_3)$ is not schedulable for any integer $a_3\geq3$.
Indeed, any occurrence of the third task after the first day would require the first task on both neighboring days, leaving three consecutive days without the second task.
These instances have density $\nicefrac{5}{6}+\nicefrac{1}{a_3}$, which approaches $\nicefrac{5}{6}$ as $a_3$ grows.
Thus, the density threshold is at most~$\nicefrac{5}{6}$.
Chan and Chin~\cite{ChanChin1993} conjectured that the density threshold is $\nicefrac{5}{6}$.
This question attracted considerable attention with many partial results obtained over the years.
Holte, Rosier, Tulchinsky, and Varvel~\cite{HOLTE1992105} showed that every integer-valued
instance with at most two distinct period values and density at most~1 is schedulable.
Lin and Lin~\cite{linlin97} proved the conjecture for integer-valued instances with at most three distinct period values.
Important intermediate results include the $\nicefrac{3}{4}$-bound of Fishburn and Lagarias~\cite{FishburnLagarias2002}, Ding's $\nicefrac{5}{6}$-bound for dimension at most~$5$~\cite{Ding2019,Ding2020}, and the $\nicefrac{5}{6}$-bound for dimension at most~$12$ by G\k{a}sieniec, Smith, and Wild~\cite{Gasieniec2022}.
Kawamura settled the conjecture of Chan and Chin in the affirmative at STOC 2024~\cite{Kawamura2024}; a full version appeared in 2026~\cite{Kawamura2026} with the following theorem.
\begin{theorem}[Kawamura~\cite{Kawamura2026}]
  \label{thm:integer}
  If an integer-valued instance $A$
  satisfies $\dens(A) \leq \nicefrac{5}{6}$, then $A$ is schedulable.
\end{theorem}

Kawamura's proof~\cite{Kawamura2024,Kawamura2026} is computer-assisted.
The reduction repeatedly combines the two tasks with largest periods into a single task whose period is half the smaller original period; a schedule for the new instance can be converted into one for the original instance by alternating between the two tasks.
Tracking the potential controls the increase in density.
A rounding step reduces the remaining verification to finitely many bounded-size integer-valued instances whose schedulability is verified by a computer program.
The intermediate periods may be nonintegral, which led Kawamura to introduce the following real-period model.
In this extended model, a real-valued instance is a tuple $A=(a_1,\ldots,a_n)\in\mathbb R^n$ with $1\leq a_1\leq\cdots\leq a_n$.
A schedule $S$ is valid if, for each task $i\in[n]$, it satisfies the following condition:
\begin{quote}
  for each $\ell \in \N$ and $m \in \N$,
  there exist at least $\ell$ values of $t \in [m, m + \lceil \ell
  a_{i} \rceil) \cap \N$ such that $S(t) = i$.
\end{quote}

Equivalently, for every positive integer $\ell$, task $i$ is performed at least $\ell$ times in any $\lceil \ell a_i\rceil$ consecutive days.
For example, when $a_i=\nicefrac{7}{2}$, task $i$ must be performed at least once in every four consecutive days, twice in every seven, three times in every eleven, and so on.
Repeating two consecutive executions followed by five days without the task satisfies the twice-in-seven requirement but violates the once-in-four requirement.
For integer periods, this agrees with the classical condition.

Kawamura~\cite{Kawamura2026} proved that any real-valued instance with at most two distinct periods and density at most~1 is schedulable.
For two tasks with irrational periods and density~1, schedulability follows from Rayleigh's theorem~\cite{Rayleigh1894}, also known as Beatty's theorem, on complementary sequences.
Fujiwara, Miyagi, and Ouchi~\cite{FujiwaraMiyagiOuchi2026} showed that any real-valued instance with at most three distinct period values and density at most $\nicefrac{5}{6}$ is schedulable and conjectured that the same bound holds for all real-valued instances.

Kawamura~\cite[Conj.~1]{Kawamura2025} conjectured a density bound depending on the minimum period:
\begin{conjecture}[Kawamura~\cite{Kawamura2025}]
  \label{conj:real_a_1}
  If an integer-valued instance $A = (a_1,a_2,\ldots,a_n)$ satisfies
  $\dens(A) \leq 1 - \frac{a_{1} - 1}{a_{1}(a_{1} + 1)}$, then $A$ is
  schedulable.
\end{conjecture}

This bound equals $\nicefrac{5}{6}$ for $a_1=2$ or $3$, is larger for $a_1\geq4$, and tends to~1 as $a_1$ grows.
Mishra, Rho, and Kleinberg~\cite{MishraRhoKleinberg2025} proved that every integer-valued instance with $a_1\geq4$ and density at most $0.84$ is schedulable.

Our main theorem establishes the conjecture of Fujiwara, Miyagi, and
Ouchi~\cite{FujiwaraMiyagiOuchi2026} and additionally bounds the
period lengths of the schedules.
\begin{theorem}
  If a real-valued instance $A$ of dimension $n$
  satisfies $\dens(A) \leq \nicefrac{5}{6}$, then $A$ admits a valid periodic schedule
  of period length less than $4\cdot2^n$.
  \label{thm:real}
\end{theorem}
For minimum period at least $4$, our approach yields the following stronger density bound.
\begin{theorem}
  \label{thm:real4}
  If a real-valued instance $A=(a_1,\ldots,a_n)$ satisfies
  $a_1\geq4$ and $\dens(A)\leq\nicefrac{17}{20}$, then $A$ admits a valid periodic schedule
  of period length less than $19\cdot2^n$.
\end{theorem}
Since $\frac{17}{20}=1-\frac{4-1}{4(4+1)}=0.85$, this improves the bound of
Mishra, Rho, and Kleinberg and establishes the $a_1=4$
case of \cref{conj:real_a_1}, even for real periods.
The bound $\nicefrac{17}{20}$ is tight: Kawamura~\cite{Kawamura2025} observed that the instances $(4,5,5,5,a_5)$ with $a_5\geq5$ are not schedulable.
Their density is $\nicefrac{17}{20}+\frac{1}{a_5}$, which approaches $\nicefrac{17}{20}$ as $a_5$ grows.

Our proof of \cref{thm:real} constructs a valid periodic schedule for every instance satisfying its assumptions.
A schedule is \emph{periodic} if there is a positive integer $L$ such that $S(t+L)=S(t)$ for every day~$t$.
We encode tasks $1,2,3,\ldots$ by the capital letters $\code{A},\code{B},\code{C},\ldots$, respectively.
For $n\in\N$, an \emph{$n$-word} is a finite word that contains each of the first $n$ letters of the alphabet at least once and no other letters.
An $n$-word $w=w_1\cdots w_L$ defines a periodic schedule $S_w$ for
$n$ tasks by repeating $w$ indefinitely.  For example, the $3$-word $w=\code{AABACB}$
represents the schedule~$S_w$ that assigns tasks $1,1,2,1,3,2$ to six
consecutive days and repeats this assignment indefinitely.

The following three lemmas establish schedulability under the density assumption of \cref{thm:real}.
The first two lemmas handle instances of dimension at most $9$.
The third lemma reduces instances of dimension at least $9$ to instances of dimension $9$.
\begin{restatable}{lemma}{smallDimensionsLemma}
  If a real-valued instance $A$ has dimension at most $8$ and satisfies $\dens(A) \leq \nicefrac{5}{6}$, then $A$ is schedulable.
  \label{lem:real8}
\end{restatable}
\begin{restatable}{lemma}{nineTaskLemma}
  If a real-valued instance $A$ has dimension $9$ and satisfies $\pot(A) \leq \nicefrac{5}{6}$, then $A$ is schedulable.
  \label{lem:real9}
\end{restatable}
\begin{restatable}{lemma}{reductionLemma}
  If a real-valued instance $A$ has dimension at least $9$ and
  satisfies $\pot(A) \leq \nicefrac{5}{6}$, then there exists a real-valued
  instance $B$ of dimension~$9$ and $\pot(B) \leq \nicefrac{5}{6}$ such that
  if $B$ is schedulable, then $A$ is schedulable.
  \label{lem:reduction}
\end{restatable}
Together, \cref{lem:real8,lem:real9,lem:reduction} show that every real-valued instance $A$ with $\dens(A)\leq\nicefrac{5}{6}$ is schedulable.
If $A$ has dimension at most $8$, apply \cref{lem:real8}.
Otherwise, $\pot(A)\leq\dens(A)\leq\nicefrac{5}{6}$, so \cref{lem:reduction} reduces the problem to an instance of dimension~$9$ to which \cref{lem:real9} applies.
The potential bound is a weaker assumption than the density bound, so \cref{lem:real9} is stronger than the corresponding density statement in dimension~$9$.
In \cref{sec:results}, we complete the proof of \cref{thm:real} by constructing periodic schedules from the certificates and the unfolding procedure, and bounding their period lengths.

We say that a schedule \emph{dominates} an instance if it is valid for that instance.
An $n$-word $w$ dominates an instance of dimension $n$ if $S_w$ is valid for that instance.
We use the same terminology for the density vector of an instance.
The proofs of \cref{lem:real8,lem:real9} are computer-assisted and verify that finitely many schedules dominate all instances in the respective domains using nine satisfiability modulo theories (SMT) queries over linear real arithmetic.
The crucial observation is that increasing a task's period relaxes its requirements.
The integer-period version of the following monotonicity lemma was already stated by Chan and Chin~\cite{ChanChin1993}.
\begin{lemma}[Kawamura~\cite{Kawamura2026}, Chan and Chin~\cite{ChanChin1993}]
  If a schedule $S$ is valid for a real-valued instance $(a_1,a_2,\ldots,a_n)$, then $S$ is also valid for the instance obtained by
  replacing some period~$a_{j}$ ($j \in [n]$) by a real number $b \geq
  a_{j}$.
  \label{lem:monotonicity}
\end{lemma}
If the periods are subsequently reordered, the task labels in the schedule are permuted accordingly.
By \cref{lem:monotonicity}, if a schedule is valid for an instance with density vector $(e_1,\dots,e_n)$, it is also valid for any instance whose density vector $(d_1,\dots,d_n)$ satisfies $d_i\leq e_i$ for every~$i$.
A family of schedules or words dominates a domain of instances if each instance in the domain is dominated by at least one member of the family.
Thus, a finite family may dominate an entire domain.
We give two examples.

\begin{figure}[tb]
  \centering
  \begin{minipage}[c]{.53\textwidth}
    \includegraphics{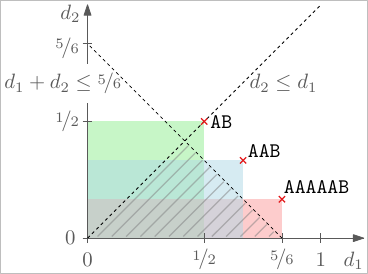}
  \end{minipage}
  \hfill
  \begin{minipage}[c]{.43\textwidth}
    \centering
    \definecolor{dimtwogreen}{rgb}{0.565,0.933,0.565}
    \definecolor{dimtwoblue}{rgb}{0.678,0.847,0.902}
    \definecolor{dimtwored}{rgb}{0.984,0.603,0.6}
    \renewcommand{\arraystretch}{1.25}
    \begin{tabular}{@{}lc@{}}
      \toprule
      Word & Density bounds\\
      \midrule
      \colorbox{dimtwogreen!50!white}{\code{AB}} & $(\nicefrac{1}{2},\nicefrac{1}{2})$\\
      \colorbox{dimtwoblue!50!white}{\code{AAB}} & $(\nicefrac{2}{3},\nicefrac{1}{3})$\\
      \colorbox{dimtwored!50!white}{\code{AAAAAB}} & $(\nicefrac{5}{6},\nicefrac{1}{6})$\\
      \bottomrule
    \end{tabular}
  \end{minipage}
  \caption{Domination in dimension~$2$ by three periodic schedules. Hatching marks the
    domain to be dominated, and each colored rectangle consists of
    the density vectors dominated by its schedule. The table lists the
    coordinates of the points marked with red crosses in the
    $(d_1,d_2)$-plane.}
  \label{fig:dim2}
\end{figure}

For dimension~$2$, it is easy to see that the domain of density vectors
$(d_1,d_2)$ with $d_1+d_2 \leq \nicefrac{5}{6}$ and $0 < d_2 \leq d_1$ is
dominated by the $2$-words \code{AAAAAB}, \code{AAB}, and \code{AB}, whose schedules are
valid for the density vectors $(\nicefrac{5}{6},\nicefrac{1}{6})$, $(\nicefrac{2}{3},\nicefrac{1}{3})$, and
$(\nicefrac{1}{2},\nicefrac{1}{2})$, respectively; see \cref{fig:dim2}.

\begin{figure}[tbp]
  \centering
  \includegraphics{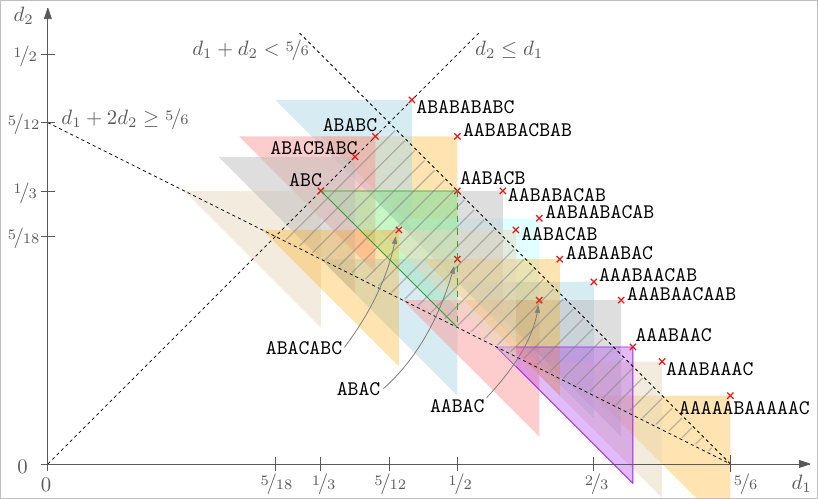}
  \par\medskip
  \begin{minipage}[c]{.62\textwidth}
    \centering
    \includegraphics{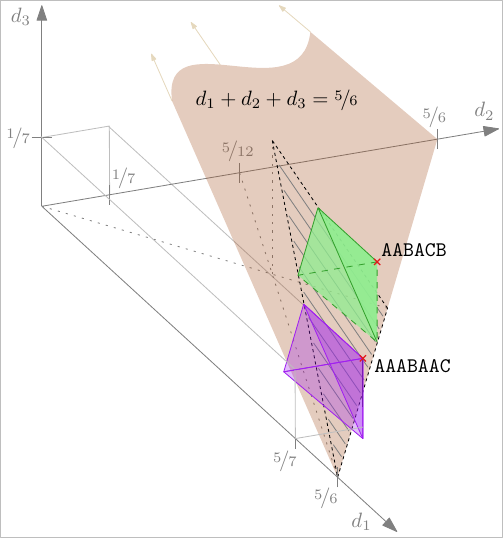}
  \end{minipage}
  \hfill
  \begin{minipage}[c]{.34\textwidth}
    \centering
    \definecolor{mylightgreen}{RGB}{198,246,202}
    \definecolor{mylightpurple}{RGB}{226,188,249}
    \begin{tabular}{@{}l@{\quad}l@{}}
      \toprule
      Word & Density bounds\\
      \midrule
      \code{AAAAABAAAAAC} & $(\nicefrac{5}{6},\nicefrac{1}{12},\nicefrac{1}{12})$\\
      \code{AAABAAAC} & $(\nicefrac{3}{4},\nicefrac{1}{8},\nicefrac{1}{8})$\\
      \colorbox{mylightpurple}{\code{AAABAAC}} & $(\nicefrac{5}{7},\nicefrac{1}{7},\nicefrac{1}{7})$\\
      \code{AAABAACAAB} & $(\nicefrac{7}{10},\nicefrac{1}{5}{-}\varepsilon,\nicefrac{1}{10})$\\
      \code{AAABAACAB} & $(\nicefrac{2}{3}{-}\varepsilon,\nicefrac{2}{9},\nicefrac{1}{9})$\\
      \code{AABAABAC} & $(\nicefrac{5}{8},\nicefrac{1}{4}{-}\varepsilon,\nicefrac{1}{8})$\\
      \code{AABAABACAB} & $(\nicefrac{3}{5}{-}\varepsilon,\nicefrac{3}{10},\nicefrac{1}{10})$\\
      \code{AABABACAB} & $(\nicefrac{5}{9},\nicefrac{1}{3}{-}\varepsilon,\nicefrac{1}{9})$\\
      \code{AABABACBAB} & $(\nicefrac{1}{2}{-}\varepsilon,\nicefrac{2}{5},\nicefrac{1}{10})$\\
      \code{AABAC} & $(\nicefrac{3}{5},\nicefrac{1}{5},\nicefrac{1}{5})$\\
      \code{AABACAB} & $(\nicefrac{4}{7},\nicefrac{2}{7},\nicefrac{1}{7})$\\
      \colorbox{mylightgreen}{\code{AABACB}} & $(\nicefrac{1}{2}{-}\varepsilon,\nicefrac{1}{3},\nicefrac{1}{6})$\\
      \code{ABABABABC} & $(\nicefrac{4}{9},\nicefrac{4}{9},\nicefrac{1}{9})$\\
      \code{ABABC} & $(\nicefrac{2}{5},\nicefrac{2}{5},\nicefrac{1}{5})$\\
      \code{ABAC} & $(\nicefrac{1}{2},\nicefrac{1}{4},\nicefrac{1}{4})$\\
      \code{ABACABC} & $(\nicefrac{3}{7},\nicefrac{2}{7},\nicefrac{2}{7})$\\
      \code{ABACBABC} & $(\nicefrac{3}{8},\nicefrac{3}{8},\nicefrac{1}{4})$\\
      \code{ABC} & $(\nicefrac{1}{3},\nicefrac{1}{3},\nicefrac{1}{3})$\\
      \bottomrule
    \end{tabular}
  \end{minipage}
  \caption{Domination in dimension~$3$ by 18 periodic schedules.
    Left: 3D view of two of the 18 schedules and the density regions
    that they dominate above the plane
    $\pi \colon d_1+d_2+d_3=\nicefrac{5}{6}$ (brown).  Top: triangular
    regions that the 18 schedules dominate on~$\pi$, vertically
    projected onto the $(d_1,d_2)$-plane.
    Dashed green edges mark the excluded face $d_1=\nicefrac{1}{2}$
    for \code{AABACB}; this face projects to a single edge in the top view.
    Hatching marks the domain to be dominated.  Each word labels the
    upper bounds on $(d_1,d_2)$ for its schedule.  The table lists the
    full upper-bound vectors; for a density~$d_i$, the value
    $x-\varepsilon$ denotes a strict bound $d_i<x$, so a marked upper
    bound need not itself be attained.}
  \label{fig:dim3}
\end{figure}

For dimension~$3$, monotonicity implies that it suffices to dominate
instances with $d_1+d_2+d_3=\nicefrac{5}{6}$ and
$0<d_3\leq d_2\leq d_1$.
Each $3$-word determines a vector of upper bounds on the three task densities, represented by a point in the $(d_1,d_2,d_3)$-coordinate system.
Some bounds may be strict, so this point need not itself be dominated.
The region dominated on the plane $\pi\colon d_1+d_2+d_3=\nicefrac{5}{6}$ consists of the positive density vectors on $\pi$ satisfying these coordinate bounds; see \cref{fig:dim3} (lower left).
\Cref{fig:dim3} (top) shows these regions projected vertically onto
the $(d_1,d_2)$-plane.
The projected domain that needs to be dominated is defined by $0<d_2\leq d_1$, $d_1+d_2<\nicefrac{5}{6}$, and $d_1+2d_2\geq\nicefrac{5}{6}$.
Each marked point in the figures is labeled by the word defining its schedule and gives the upper bounds on the first two task densities.
For example, \code{AAABAAC} (light purple in \Cref{fig:dim3})
dominates all density vectors with $d_1\leq\nicefrac{5}{7}$,
$d_2\leq\nicefrac{1}{7}$, and $d_3\leq\nicefrac{1}{7}$.  Some bounds
are strict: \code{AABACB} (light green in \Cref{fig:dim3}) dominates
exactly the positive density vectors with $d_1<\nicefrac{1}{2}$,
$d_2\leq\nicefrac{1}{3}$, and $d_3\leq\nicefrac{1}{6}$.
The strict bound on $d_1$ is necessary because the consecutive letters \code{CB} leave two days without task~$1$.
We explain how to compute all these bounds in \cref{sec:domains}.

The two-task domination argument and the use of density coordinates and projection in the three-task case already appear in Fujiwara, Miyagi, and Ouchi~\cite{FujiwaraMiyagiOuchi2026,FujiwaraMiyagiOuchi-SOFSEM2026}.
These examples were the starting point for our project on automated proofs of analogous finite domination results in higher dimensions.
Their hand-crafted proof for three tasks divides the domain into seven
regions, using six fixed schedules and an instance-dependent
construction.  Our computer-generated certificate consists of 18 fixed
periodic schedules that dominate the entire domain.

The proof of \cref{lem:reduction} uses \emph{folding}: the two tasks
with the two largest periods, $a_{n-1}$ and $a_n$, where $a_{n-1}\leq a_n$,
are replaced by a single task of period ${a_{n-1}}/{2}$.
A valid schedule for the folded instance yields one for the original instance by assigning successive occurrences of the new task alternately to the two original tasks.
This uses the following unfolding principle.

\begin{lemma}[Kawamura~\cite{Kawamura2026}, G\k{a}sieniec, Smith, and Wild~\cite{Gasieniec2022}]
  If a real-valued instance $A=(a_1,\ldots,a_n)$ is schedulable,
  then replacing a task of period $a_j$ by $q$ tasks of period
  $qa_j$, for any $q\in\N$, and reordering the periods
  produces a schedulable instance.
  \label{lem:partitioning}
\end{lemma}

Given a valid schedule for the folded instance, we apply the lemma with $q=2$
to the task of period ${a_{n-1}}/{2}$, obtaining two tasks of period $a_{n-1}$.
We then use monotonicity to increase one of these periods to $a_n$.
Folding and unfolding for integer-valued instances were explicitly defined by G\k{a}sieniec, Smith, and Wild~\cite{Gasieniec2022}.
Additionally, folding the two tasks with largest periods does not increase the potential.
Repeating this operation therefore reduces the dimension to~$9$ while maintaining the potential bound.

For the $\nicefrac{17}{20}$-bound in \cref{thm:real4}, folding must also preserve the minimum period~$4$.
We stop the reduction at ten tasks, use a restricted nine-task potential certificate for cases reduced by one further fold, and verify the remaining ten-task cases directly; see \cref{sec:periods-at-least-4}.

\paragraph*{Related work.}

The integer-valued pinwheel scheduling problem was introduced by
Holte, Mok, Rosier, Tulchinsky, and Varvel~\cite{Holte1989}.  They
also showed that the corresponding decision problem is in \PSPACE.
Kleinberg and Mishra~\cite{KleinbergMishra2026} recently showed that
it is \NP-hard to decide whether a given integer-valued pinwheel
scheduling instance is schedulable.
They also gave an efficient polynomial-time approximation scheme for the approximate decision problem: for an integer-valued instance $A$ and any $\varepsilon>0$, it either certifies that $A$ is unschedulable or that $((1+\varepsilon)a_1,(1+\varepsilon)a_2,\ldots,(1+\varepsilon)a_n)$ is schedulable in the real-period model.
Kobayashi, Lin, and
Swernofsky~\cite{KobayashiLinSwernofsky2026} proved that deciding schedulability
for integer-valued pinwheel instances of density exactly~1 is strongly
\NP-complete.

G\k{a}sieniec, Smith, and Wild~\cite{Gasieniec2022} showed that integer-valued pinwheel scheduling is fixed-parameter tractable with respect to the number of tasks.
Kobayashi and Lin~\cite{KobayashiLin2025}, with a journal extension by Kobayashi, Lin, and Swernofsky~\cite{KOBAYASHI2026115998}, gave explicit running-time bounds and established fixed-parameter tractability for the covering and exact-period variants as well.

The pinwheel scheduling problem is one of the fundamental problems in
\emph{perpetual scheduling}, in which the output is a schedule of
infinite length.  The Bamboo Garden Trimming Problem (BGT) is another
example of such a
problem~\cite{Kuszmaul2022,GASIENIEC2024103476,Kawamura2026}.

The pinwheel scheduling problem can be viewed as a task packing
problem.  Its dual problem, the \emph{pinwheel covering
problem}~\cite{KAWAMURA2020195}, is defined as follows: Assuming that
each agent $i \in [n]$ can work at most once in any $a_{i}$
consecutive days, we are asked to find a schedule that assigns at least
one agent to each day.
The optimal density threshold for integer-valued pinwheel covering is
\[
  \alpha^*=\sum_{j=0}^{\infty}\frac{1}{2^j+1}\approx1.264\text{,}
\]
i.e., every pinwheel covering instance of density at least $\alpha^*$ is schedulable.
This was proved independently by Mishra~\cite{Mishra2026} and by Kawamura and Kobayashi~\cite{KawamuraKobayashi2026} using computer assistance.

\paragraph*{AI-assisted proofs and attribution.}

In September 2026, we received communications about AI-assisted proofs of the real-period $\nicefrac{5}{6}$-bound, including an unpublished proof by the anonymous Nuntiatio Collective~\cite{Nuntiatio2026} and a publicly available manuscript by Fetik~\cite{Fetik2026}.
Later, we received further claims of AI-generated proofs without details of the arguments.

By then, we had already implemented a framework for automated proof generation and verification and verified the density bound $\nicefrac{5}{6}$ for instances of dimension up to~$10$.
This earlier verification used roughly half a million explicit schedules to dominate all instances in that domain.
The reductions in the communicated AI-assisted proofs drew our attention to the potential function, which
was introduced by Kawamura (\cite[Lemma~4]{Kawamura2024}, \cite[Lemma~5]{Kawamura2026}).
Incorporating the potential into our framework yielded a compact proof based on
finite families of periodic schedules and SMT verification of domination.

\paragraph*{Organization.}
We prove the dimension-reduction lemma in \cref{sec:folding} and introduce the density domains and dominance boxes in \cref{sec:domains}.
\Cref{sec:results} verifies domination by finite families of schedules and derives an explicit bound on the lengths of the periods of these schedules.
\Cref{sec:periods-at-least-4} establishes the tight density bound
$\nicefrac{17}{20}$ for real-valued instances with minimum period at
least~$4$.

\section{Dimension Reduction}
\label{sec:folding}

\reductionLemma*

\begin{proof}
For $n=9$, take $B=A$.
Let $A=(a_1, a_2, \ldots, a_n)$ have dimension $n>9$ and satisfy $\pot(A)\leq\nicefrac{5}{6}$.
Let $(x,y,z)=(d_{n-2},d_{n-1},d_n)$.  Then $x \geq y \geq z$.
Fold the last two tasks with periods $a_{n-1}$ and $a_n$ into a single
task of period ${a_{n-1}}/{2}$ and density $2y$, and reorder the resulting
periods to obtain an instance $A'$ of dimension $n-1$.
Since $(n-1)y\leq\pot(A)\leq\nicefrac{5}{6}$ and $n\geq10$, the folded density satisfies $2y\leq\nicefrac{5}{27}<1$, so the new period is at least~$1$.
We have $\dens(A')=\dens(A)+y-z$, and the smallest task density of $A'$ is $\min\{x,2y\}$.  Consequently,
\begin{align}
 \pot(A')
 &=\dens(A')-\min\{x,2y\}\notag\\
 &=\dens(A)+y-z-\min\{x,2y\}\notag\\
 &=\pot(A)+y-\min\{x,2y\}\notag\\
 &=\pot(A)-\min\{x-y,y\}.
 \label{eq:potential-change}
\end{align}
Thus, $\pot(A') \leq \pot(A) \leq \nicefrac{5}{6}$.
The unfolding argument of \cref{lem:partitioning} shows that the schedulability of $A'$ implies the schedulability of $A$.
Repeating the fold until dimension $9$ is reached yields the required instance $B$.
\end{proof}

Observe that the total density may increase during a fold,
whereas the potential does not.
This explains why our approach uses the potential bound in \cref{lem:real9}.

The unfolding construction also preserves periodicity.
Let an $n$-word $w$ define a valid schedule for the folded instance.
If the folded task occurs an even number of times in $w$, replace its occurrences alternately by the two original tasks; if it occurs an odd number of times, do this in $ww$ instead.
After relabeling the tasks according to their periods, the result is an $(n+1)$-word defining a valid schedule for the original instance and having length at most $2|w|$.
Thus, each unfolding increases the word length by a factor of at most~2.

\section{Density Domains and Dominance Boxes}
\label{sec:domains}

It remains to prove \cref{lem:real8,lem:real9}.
We use a finite family of periodic schedules for each of the required dimensions.
Writing $\densities=(d_1,\ldots,d_n)$ for a density vector, define
\[
 \mathcal D_n=\left\{\densities\in\mathbb R^n:
 d_1\geq\cdots\geq d_n>0,\quad\textstyle{\sum_{i=1}^n} d_i\leq\nicefrac{5}{6}\right\}\text{,}
\]
and
\[
 \mathcal P_n=\left\{\densities\in\mathbb R^n:
 d_1\geq\cdots\geq d_n>0,\quad\textstyle{\sum_{i=1}^{n-1}} d_i\leq\nicefrac{5}{6}\right\}\text{.}
\]
Domains $\mathcal D_1,\ldots,\mathcal D_8$ correspond to \cref{lem:real8},
and domain~$\mathcal P_9$ corresponds to \cref{lem:real9}.
Every vector in these domains has $d_1\leq\nicefrac{5}{6}$, so its periods $a_i=\frac{1}{d_i}$ satisfy $a_i\geq1$.
For an instance $A$ whose density vector belongs to $\mathcal P_9$, the ordering of the densities gives a bound on the total density:
\[
 d_9 \leq {\pot(A)}/{8},
 \qquad
 \dens(A)=\pot(A)+d_9
 \leq \nicefrac{9}{8} \cdot \pot(A) \leq \nicefrac{15}{16}.
\]
We first describe the dominance box of a periodic schedule and illustrate it with an example.
We then prove a formula for the intervals defining this box.
For an $n$-word $w=w_1\cdots w_L$, recall that $S_w$ is the periodic schedule obtained by repeating $w$ indefinitely.
We keep task labels fixed and allow positive task densities in arbitrary order.
The task requirements are independent, and validity is monotone in each task density.
Hence, the density vectors for which $S_w$ is valid form a product of $n$ intervals.
We call this product the \emph{dominance box} $B(w)$ of~$w$ and write
\[
 B(w)=I_1(w)\times\cdots\times I_n(w),
 \qquad I_i(w)=(0,u_i)\text{ or }(0,u_i],
\]
where $I_i(w)$ is the interval of admissible densities for task $i$ and $u_i>0$ is its upper endpoint.
An $n$-word $w$ is \emph{density-ordered} if
\[
 I_1(w)\supseteq I_2(w)\supseteq\cdots\supseteq I_n(w).
\]
Thus, the upper endpoints are non-increasing; when two endpoints coincide, a closed interval precedes an open one.
Every $n$-word can be made density-ordered by relabeling its letters.
Moreover, this relabeling preserves domination of every non-increasing density vector already dominated by the word.
Indeed, if $I_i(w)\subsetneq I_{i+1}(w)$, swapping the letters encoding tasks $i$ and $i+1$ preserves domination: $d_i\in I_i(w)\subseteq I_{i+1}(w)$, and $0<d_{i+1}\leq d_i$ implies $d_{i+1}\in I_i(w)$.
Repeating such swaps orders the intervals, so restricting attention to density-ordered words loses no dominated instances.
Even for a density-ordered word, the points of $B(w)$ need not have non-increasing coordinates; our instance domains still require $d_1\geq\cdots\geq d_n>0$.

The introductory example $w=\code{AABACB}$ is density-ordered, with
\[
 B(w)=\left(0,\nicefrac{1}{2}\right)\times\left(0,\nicefrac{1}{3}\right]\times\left(0,\nicefrac{1}{6}\right].
\]
The first bound is strict even though \code{A} occupies half the days.
At $d_1=\nicefrac{1}{2}$, task $1$ would have period $a_1=2$ and would have to occur in every two consecutive days.
The consecutive letters \code{CB} in $w$ violate that requirement.
The bounds for \code{B} and \code{C} include their endpoints: these tasks occur every three and every six days, respectively.
The certificate records the word and these bounds as
\begin{center}
\code{AABACB,<1/2,<=1/3,<=1/6}
\end{center}
Our C++ program \code{packing\_schedule\_quality} computes the dominance box of a given $n$-word~$w$ and is available in the accompanying repository~\cite{Software}.
In the program, we adapt the schedule-domination approach of G\k{a}sieniec, Smith, and Wild~\cite{Gasieniec2022} to real periods.
(Their accompanying programs are available
online.\footnote{\href{https://github.com/roarin-roran/Towards-the-5-over-6-Density-Conjecture-of-Pinwheel-Scheduling}{Source code accompanying the ALENEX 2022 paper.}})
For integer periods, determining which instances a schedule dominates
requires only the largest cyclic gap between successive occurrences of
each task.
For real periods, the calculation is similar, but we also have to
consider spans that contain several consecutive gaps.
Two copies of $w$, of total length $2L$, suffice to evaluate all cyclic
spans as we show~now.

For each task $i\in[n]$, let $m_i>0$ be its number of occurrences in $w$,
and, for $k\in\N$, let $G_i(k)$ be the maximum distance in days from
an occurrence of task~$i$ to the $k$-th following occurrence in $S_w$,
allowing spans across period boundaries.
Adapting the periodicity argument of Fujiwara, Miyagi, and
Ouchi~\cite{FujiwaraMiyagiOuchi2026} to Kawamura's density
formulation~\cite{Kawamura2026} yields the upper endpoint~$u_i$ of~$I_i(w)$:
\begin{equation}\label{eq:quality}
 u_i=\min\left\{
 \frac{m_i}{L},\,
 \frac{1}{G_i(1)-1},\,\ldots,\,
 \frac{m_i}{G_i(m_i)-1}
 \right\}.
\end{equation}
Gap terms with $G_i(k)=1$ are omitted.
The term ${m_i}/{L}$ is the mean bound, and the quotients $\frac{k}{G_i(k)-1}$ are gap bounds.
The endpoint $u_i$ is excluded if any gap bound attains this minimum; otherwise, it is included.

For $w=\code{AABACB}$, we have $L=6$ and $m_1=3$, with
$G_1(1)=3$, $G_1(2)=5$, and $G_1(3)=6$.
Thus,
\[
 u_1=\min\left\{\nicefrac{3}{6},\nicefrac{1}{2},\nicefrac{2}{4},\nicefrac{3}{5}\right\}
 =\nicefrac{1}{2}.
\]
The gap bounds for $k=1$ and $k=2$ attain this minimum, so $I_1(w)=(0,\nicefrac{1}{2})$.

\begin{proof}[Proof of~\eqref{eq:quality} and the endpoint rule]
For positive integers $h,k$, every block of $h$ days contains at
least $k$ occurrences of task $i$ if and only if $G_i(k)\leq h$.
Indeed, shifting a deficient block forward by $L$ days preserves its task counts and ensures that a preceding occurrence of task~$i$ exists.
The shifted block lies strictly between the last such occurrence and its $k$-th successor, so $G_i(k)>h$.
Conversely, a span longer than $h$ contains a deficient $h$-day block.
Thus, validity at density $d_i>0$ is equivalent~to
\[
 G_i(k) \leq \left\lceil \frac{k}{d_i} \right\rceil
 \quad\Longleftrightarrow\quad
 d_i \cdot \bigl(G_i(k)-1\bigr)<k
 \qquad(k\geq1).
\]
When $G_i(k)=1$, this condition is automatic; otherwise, it gives
the strict gap bound $d_i<\frac{k}{G_i(k)-1}$.
Since $G_i(qm_i)=qL$, letting $q$ tend to infinity yields
$d_iL\leq m_i$, the closed mean bound.
Conversely, suppose the mean bound and the gap bounds for
$1\leq k\leq m_i$ hold.
Write any positive integer $k$ as $k=qm_i+r$, where $q\geq0$
and $1\leq r\leq m_i$.
Periodicity gives $G_i(k)=qL+G_i(r)$, and hence
\[
 d_i \cdot \bigl(G_i(k)-1\bigr)
 =q\,d_iL+d_i \cdot \bigl(G_i(r)-1\bigr)
 <qm_i+r=k.
\]
Thus, these finitely many bounds characterize $I_i(w)$, yielding~\eqref{eq:quality} and the stated endpoint rule.
\end{proof}
Thus, an $n$-word $w$ dominates an instance $A$ of dimension $n$ if and only if the density vector of $A$ belongs to $B(w)$.

For the proofs of \cref{lem:real8,lem:real9}, we define a finite set $\mathcal W_n$ of density-ordered $n$-words for each $n\in\{1,\ldots,9\}$.
It remains to verify that the respective domains are contained in the unions of the dominance boxes defined by these words:
\begin{equation}\label{eq:cover}
 \mathcal D_n\subseteq\bigcup_{w\in\mathcal W_n}B(w)
 \quad(1\leq n\leq8),
 \qquad
 \mathcal P_9\subseteq\bigcup_{w\in\mathcal W_9}B(w).
\end{equation}
G\k{a}sieniec, Smith, and Wild~\cite{Gasieniec2022} used finite Pareto surfaces to verify the $\nicefrac{5}{6}$-bound for integer-valued instances with at most 12 tasks.
Here we verify that finite families of periodic schedules dominate entire domains of real density vectors.

The domains and boxes in~\eqref{eq:cover} are defined by linear inequalities with rational coefficients.
For finite systems of such constraints, including strict inequalities, satisfiability over $\mathbb Q$ is equivalent to satisfiability over $\mathbb R$~\cite[Section~4]{BotteschHaslbeckThiemann2019}.
Applying this result to the finite Boolean combination describing density vectors outside all dominance boxes shows that~\eqref{eq:cover} holds for real densities if and only if it holds for rational densities.
Our verification uses real variables and exact rational coefficients.
Moreover, it retains strict endpoint conditions and includes the
$\nicefrac{5}{6}$-boundary in terms of density or potential.

\section{The Computer-Assisted Proofs}
\label{sec:results}
\label{sec:implementation}

We verify the inclusions in~\eqref{eq:cover}, proving \cref{lem:real8,lem:real9}, and then bound the period length of the schedules obtained by unfolding.
We begin with the two-task example from \cref{fig:dim2}.
The $2$-words \code{AAAAAB}, \code{AAB}, and \code{AB} have dominance boxes
$(0,\nicefrac{5}{6}] \times (0,\nicefrac{1}{6}]$, $(0,\nicefrac{2}{3}] \times (0,\nicefrac{1}{3}]$, and $(0,\nicefrac{1}{2}] \times (0,\nicefrac{1}{2}]$,
respectively.
To verify domination of $\mathcal D_2$ by these words, we ask whether there is a density vector $(d_1,d_2)\in\mathcal D_2$ outside all three boxes.
The following SMT-LIB program and existential formula express this query:

\medskip

\noindent
\begingroup
\begin{tabular}{@{}p{0.52\linewidth}@{\hspace{0.03\linewidth}}p{0.45\linewidth}@{}}
\textbf{SMT-LIB program} & \textbf{Existential formula}\\[0.5ex]
\code{(set-logic QF\_LRA)} & \\
\code{(declare-const d1 Real)} & $\exists\,d_1\in\mathbb R:$\\
\code{(declare-const d2 Real)} & $\exists\,d_2\in\mathbb R:$\\
\code{(assert (and (> d2 0) (>= d1 d2)} & $\quad d_2>0\land d_1\geq d_2$\\
\hspace*{1em}\code{(<= (+ d1 d2) (/ 5 6))))} & $\quad{}\land d_1+d_2\leq \nicefrac{5}{6}$\\
\code{(assert (not (and (<= d1 (/ 5 6))} & $\quad{}\land\neg(d_1\leq \nicefrac{5}{6}$\\
\hspace*{1em}\code{(<= d2 (/ 1 6)))))} & $\qquad{}\land d_2\leq \nicefrac{1}{6})$\\
\code{(assert (not (and (<= d1 (/ 2 3))} & $\quad{}\land\neg(d_1\leq \nicefrac{2}{3}$\\
\hspace*{1em}\code{(<= d2 (/ 1 3)))))} & $\qquad{}\land d_2\leq \nicefrac{1}{3})$\\
\code{(assert (not (and (<= d1 (/ 1 2))} & $\quad{}\land\neg(d_1\leq \nicefrac{1}{2}$\\
\hspace*{1em}\code{(<= d2 (/ 1 2)))))} & $\qquad{}\land d_2\leq \nicefrac{1}{2})$\\
\code{(check-sat)} & \\
\code{(exit)} &
\end{tabular}
\endgroup
\par\medskip
The first assertion describes $\mathcal D_2$; the next three exclude the boxes of \code{AAAAAB}, \code{AAB}, and \code{AB}.
The solver returns \code{unsat}, proving that
\[
 \mathcal D_2\subseteq B(\code{AAAAAB})\cup B(\code{AAB})\cup B(\code{AB}).
\]

For the full word sets, the domination queries are
\begin{align}
 &\exists\,\densities\in\mathcal D_n:\quad
 \bigwedge_{w\in\mathcal W_n}\densities\notin B(w)
 \qquad(1\leq n\leq8),\label{eq:smt-density}\\
\intertext{and}
 &\exists\,\densities\in\mathcal P_9:\quad
 \bigwedge_{w\in\mathcal W_9}\densities\notin B(w).\label{eq:smt-potential}
\end{align}
Expanding domain and box membership gives Boolean combinations of linear inequalities with rational coefficients, including the strict bounds of open box endpoints.
The solver checks these formulas over the reals using exact arithmetic.
An unsatisfiable answer proves the corresponding inclusion in~\eqref{eq:cover}; a satisfying assignment gives a density vector not dominated by any word in the current set.

We constructed the word sets incrementally, adding a dominating word whenever a query returned an undominated rational density vector.
The words came from schedule libraries and recursive applications of the unfolding principle in \cref{lem:partitioning}.

\Cref{tab:certificates} lists the nine certificate files specifying $\mathcal W_1,\ldots,\mathcal W_9$.
They contain 6\,319 words in dimensions $1$ through $8$ and 7\,968 words for dimension $9$, for a total of 14\,287 distinct words.
Each certificate entry records a word and the coordinate bounds of its dominance box, in the format illustrated in \cref{sec:domains}.

\begin{table}[tb]
\centering
\caption{The certificates used to prove \cref{lem:real8,lem:real9}. Each density or potential bound is $\nicefrac{5}{6}$. Maximum length is the length of a longest word in each set. Files are in \code{packing\_plans/proof\_5\_6/}, relative to the software repository (the file names are clickable). Check times include dominance-box recomputation and SMT verification.}
\label{tab:certificates}
\begin{tabular}{@{}rccrrlr@{}}
\toprule
Dimension & Quantity bounded & Domain & Words & Maximum length & File & Check time\\
\midrule
1 & Density & $\mathcal D_1$ & 1 & 1 & \href{https://github.com/grzegorz-gutowski/pinwheel_scheduling/blob/main/packing_plans/proof_5_6/D1.txt}{\code{D1.txt}} & $<1$\,s\\
2 & Density & $\mathcal D_2$ & 3 & 6 & \href{https://github.com/grzegorz-gutowski/pinwheel_scheduling/blob/main/packing_plans/proof_5_6/D2.txt}{\code{D2.txt}} & $<1$\,s\\
3 & Density & $\mathcal D_3$ & 18 & 12 & \href{https://github.com/grzegorz-gutowski/pinwheel_scheduling/blob/main/packing_plans/proof_5_6/D3.txt}{\code{D3.txt}} & $<1$\,s\\
4 & Density & $\mathcal D_4$ & 69 & 18 & \href{https://github.com/grzegorz-gutowski/pinwheel_scheduling/blob/main/packing_plans/proof_5_6/D4.txt}{\code{D4.txt}} & $<1$\,s\\
5 & Density & $\mathcal D_5$ & 216 & 32 & \href{https://github.com/grzegorz-gutowski/pinwheel_scheduling/blob/main/packing_plans/proof_5_6/D5.txt}{\code{D5.txt}} & $<1$\,s\\
6 & Density & $\mathcal D_6$ & 630 & 64 & \href{https://github.com/grzegorz-gutowski/pinwheel_scheduling/blob/main/packing_plans/proof_5_6/D6.txt}{\code{D6.txt}} & 2\,s\\
7 & Density & $\mathcal D_7$ & 1\,657 & 352 & \href{https://github.com/grzegorz-gutowski/pinwheel_scheduling/blob/main/packing_plans/proof_5_6/D7.txt}{\code{D7.txt}} & 6\,s\\
8 & Density & $\mathcal D_8$ & 3\,725 & 608 & \href{https://github.com/grzegorz-gutowski/pinwheel_scheduling/blob/main/packing_plans/proof_5_6/D8.txt}{\code{D8.txt}} & 24\,s\\
9 & Potential & $\mathcal P_9$ & 7\,968 & 1\,800 & \href{https://github.com/grzegorz-gutowski/pinwheel_scheduling/blob/main/packing_plans/proof_5_6/P9.txt}{\code{P9.txt}} & 189\,s\\
\midrule
\multicolumn{3}{@{}l}{Total} & 14\,287 & & & 222\,s\\
\bottomrule
\end{tabular}
\end{table}

The repository's \code{check-proofs-5-6} command uses pySMT~\cite{GarioMicheli2015} with Z3~\cite{DeMouraBjorner2008} to check domination of all nine domains.
It recomputes the dominance box of each stored word from~\eqref{eq:quality} and its endpoint rule before checking domination, without searching for additional schedules.
The repository README file~\cite{Software} gives build and verification instructions.
Using a single core of an Apple M4 Pro CPU, checking all nine certificates takes about 4 minutes, including dominance-box recomputation and SMT verification.
\Cref{tab:certificates} gives the times for individual certificate checks.

\smallDimensionsLemma*

\nineTaskLemma*

\begin{proof}[Proof of \cref{lem:real8,lem:real9}]
For the word sets in \cref{tab:certificates}, all eight queries~\eqref{eq:smt-density} and the query~\eqref{eq:smt-potential} are unsatisfiable.
Hence the inclusions in~\eqref{eq:cover} hold.
Every instance in \cref{lem:real8} or \cref{lem:real9} is therefore dominated by a word $w$ in the corresponding set, so $S_w$ is a valid periodic schedule.
\end{proof}

The finite families of dominating words also yield a valid periodic schedule for any real-valued instance $A$ of dimension $n$ with $\dens(A)\leq\nicefrac{5}{6}$.
If $n\leq8$, choose a dominating word from the corresponding set $\mathcal W_n$.
For $n\geq9$, we have $\pot(A)\leq\dens(A)\leq\nicefrac{5}{6}$, so \cref{lem:reduction} applies.
Fold the two tasks with smallest densities repeatedly until dimension $9$ is reached, recording the two children of each folded task.
For the resulting 9-task instance, choose a dominating word from $\mathcal W_9$ and reverse the folds using the construction in \cref{sec:folding}, relabeling the tasks whenever their densities are reordered.

For rational inputs, a priority queue implements the folding steps using $O(n\log n)$ exact density comparisons; the final word set has constant size independent of $n$.
This comparison count does not include the cost of rational arithmetic or of writing the final periodic word, whose length may grow exponentially with $n$.
By \cref{sec:folding}, each unfolding increases the word length by at most a factor of two.
Together with the maximum word lengths in \cref{tab:certificates}, this gives the period-length bound in \cref{thm:real} and the following extension.
\begin{corollary}\label{cor:period}
Every real-valued instance $A$ of dimension $n\geq9$ with $\pot(A)\leq\nicefrac{5}{6}$ admits a valid periodic schedule of period length less than $4\cdot2^n$.
\end{corollary}
\begin{proof}[Proof of \cref{thm:real,cor:period}]
Let $A$ be an instance satisfying the assumptions of \cref{thm:real} or \cref{cor:period}, and let $n$ be its dimension.
Let $L_k$ be the maximum length of a word in $\mathcal W_k$, as listed in \cref{tab:certificates}.
For $1\leq k\leq9$, the table gives
\[
 \max_{1\leq k\leq9}\frac{L_k}{2^k}
 =\frac{1\,800}{2^9}=\frac{225}{64}<4.
\]
For $n\leq8$, a dominating word from $\mathcal W_n$ has length at most $L_n<4\cdot2^n$.
For $n\geq9$, either assumption on $A$ gives $\pot(A)\leq\nicefrac{5}{6}$.
The construction above uses a word from $\mathcal W_9$ and reverses $n-9$ folds, so the resulting period length is at most
$L_9\,2^{n-9}<4\cdot2^n$.
\end{proof}

\section{Periods at Least 4}
\label{sec:periods-at-least-4}

The proof of \cref{thm:real4} uses the same techniques as the proof of \cref{thm:real}: domination by finite families of periodic schedules, verified using SMT, and dimension reduction by folding.
To keep the certificate families small, we first verify schedulability on a restricted nine-task potential domain.
A single fold reduces part of the 10-task potential domain to this case; we verify the remaining part directly.
We use the following three lemmas.

\begin{restatable}{lemma}{smallDimensionsFourLemma}
  If a real-valued instance $A=(a_1,\ldots,a_n)$ has dimension at most $9$ and satisfies $a_1\geq4$ and $\dens(A) \leq \nicefrac{17}{20}$, then $A$ is schedulable.
  \label{lem:real4-small}
\end{restatable}

\begin{restatable}{lemma}{tenTaskFourLemma}
  If a real-valued instance $A=(a_1,\ldots,a_{10})$ of dimension $10$ satisfies $a_1\geq4$ and $\pot(A) \leq \nicefrac{17}{20}$, then $A$ is schedulable.
  \label{lem:real4-ten}
\end{restatable}

\begin{proof}[Proof of \cref{lem:real4-ten}]
We first claim that every real-valued nine-task instance with minimum period in $[4,\nicefrac{20}{3}]$ and potential at most $\nicefrac{17}{20}$ is schedulable.
The corresponding potential domain is
\[
 \mathcal P_9^{(4)}=\left\{\densities\in\mathbb R^9:
 \frac{1}{4}\geq d_1\geq\cdots\geq d_9>0,\quad
 d_1\geq\frac{3}{20},\quad
 \sum_{i=1}^8 d_i \leq \frac{17}{20} \right\}.
\]
The bounds $\nicefrac{3}{20}\leq d_1\leq\nicefrac{1}{4}$ express $4\leq a_1\leq\nicefrac{20}{3}$.

For the words in \code{P9.txt}, the domination query over $\mathcal P_9^{(4)}$ is unsatisfiable, as in the proof of \cref{lem:real9}.
Thus, every vector in this domain is dominated by a word $w$ in the certificate, so $S_w$ is a valid periodic schedule.
This proves the claim.

Now let $A$ satisfy the assumptions of \cref{lem:real4-ten}, with density vector $(d_1,\ldots,d_{10})$.
Folding the last two tasks replaces their densities by $2d_9$.
Since $9d_9\leq\pot(A)$, we have
\[
 2d_9 \leq \frac{2\pot(A)}{9} \leq \frac{17}{90} < \frac{1}{4}.
\]
After reordering, the resulting nine-task instance $A'$ has largest density $\max\{d_1,2d_9\}\leq\nicefrac{1}{4}$ and, by~\eqref{eq:potential-change}, potential at most $\nicefrac{17}{20}$.
If $\max\{d_1,2d_9\}\geq\nicefrac{3}{20}$, its density vector belongs to $\mathcal P_9^{(4)}$.
The claim then makes $A'$ schedulable, and unfolding yields a valid schedule for $A$.

Otherwise, $d_1<\nicefrac{3}{20}$ and $d_9<\nicefrac{3}{40}$.
These ten-task density vectors form $\mathcal P_{10}^{(4)}$:
\[
 \mathcal P_{10}^{(4)}=\left\{\densities\in\mathbb R^{10}:
 \frac{3}{20} > d_1\geq\cdots\geq d_{10}>0,\quad
 d_9 < \frac{3}{40},\quad
 \sum_{i=1}^9 d_i \leq \frac{17}{20}\right\}.
\]
On this domain, we use the same domination argument with finitely many schedules and SMT verification as in the proof of \cref{lem:real9}.
The file \code{P10.txt} certifies this residual domain; the preceding folding argument covers its complement in the full ten-task potential domain.
The boundary cases $d_1=\nicefrac{3}{20}$ or $d_9=\nicefrac{3}{40}$ are included in the folding argument.
\end{proof}

\begin{restatable}{lemma}{reductionFourLemma}
  If a real-valued instance $A=(a_1,\ldots,a_n)$ has dimension at
  least $10$ and satisfies $a_1\geq4$ and $\pot(A) \leq \nicefrac{17}{20}$, then
  there exists a real-valued instance $B=(b_1,\ldots,b_{10})$ of
  dimension $10$, $b_1\geq4$, and $\pot(B) \leq \nicefrac{17}{20}$ such that if
  $B$ is schedulable, then $A$ is schedulable.
  \label{lem:real4-reduction}
\end{restatable}

Together, \cref{lem:real4-small,lem:real4-ten,lem:real4-reduction} imply the schedulability assertion of \cref{thm:real4}.
The constructions yield valid periodic schedules, whose period lengths we bound at the end of this section.
The proof of \cref{lem:real4-small} follows that of \cref{lem:real8}, with the density bound changed to $\nicefrac{17}{20}$ and the domain restricted by $d_1\leq\nicefrac{1}{4}$.
The proof of \cref{lem:real4-reduction} follows the folding argument in \cref{lem:reduction}, stopping at dimension $10$.
At every folding step, the current dimension satisfies $n>10$, and the new task density obeys
\[
 2d_{n-1}\leq\frac{2\pot(A)}{n-1}
 \leq\frac{17}{10(n-1)}
 \leq\frac{17}{100}<\frac{1}{4}.
\]
All other task densities remain at most $\nicefrac{1}{4}$, so the folded instance still has minimum period at least $4$.
Equation~\eqref{eq:potential-change} shows that its potential remains at most $\nicefrac{17}{20}$, and unfolding preserves schedulability.
In addition to the two potential domains defined above, we use the following density domains for \cref{lem:real4-small}:
\[
 \mathcal D_n^{(4)}=\left\{\densities\in\mathbb R^n:
 \frac{1}{4} \geq d_1\geq \dots \geq d_n>0,\quad
 \sum_{i=1}^n d_i \leq \frac{17}{20}\right\}
 \qquad(1\leq n\leq9).
\]
\Cref{tab:certificates-four} lists the eleven certificate files specifying the sets used in the computer-assisted proofs.
Each file contains density-ordered words of the corresponding dimension.
The density certificates contain 11\,846 words, and the two potential certificates contain 22\,752 and 3\,747 words, respectively.
The two nine-task certificates share 6\,209 words. Counting each word only once across all eleven files gives 32\,136 distinct words.

The repository command \code{check-proofs-17-20} verifies all eleven certificate domains in \cref{tab:certificates-four}; all eleven SMT queries are unsatisfiable.
Using a single core of an Apple M4 Pro CPU, checking all eleven certificates takes about 21 minutes, including dominance-box recomputation and SMT verification.
\Cref{tab:certificates-four} gives the times for individual certificate checks.

\begin{proof}[Proof of \cref{thm:real4}]
Let $A$ be an instance satisfying the assumptions of \cref{thm:real4}, and let $n$ be its dimension.
For $1\leq k\leq9$, let $L_k^{(4)}$ be the maximum word length in the certificate for $\mathcal D_k^{(4)}$.
\Cref{tab:certificates-four} gives
\[
 \max_{1\leq k\leq9}\frac{L_k^{(4)}}{2^k}
 =\frac{4\,752}{2^8}=\frac{297}{16}<19.
\]
Thus, for $n\leq9$, a dominating word has length at most $L_n^{(4)}<19\cdot2^n$.
For $n\geq10$, we have $\pot(A)\leq\dens(A)\leq\nicefrac{17}{20}$, so \cref{lem:real4-reduction} reduces $A$ to dimension $10$.
The proof of \cref{lem:real4-ten} then uses either a word of length at most $3\,348$ from the ten-task certificate, or one further fold followed by a word of length at most $3\,960$ from the nine-task potential certificate.
Since each unfolding increases the period length by a factor of at most $2$, the resulting schedule has period length at most
\[
 \max\{3\,348,2\cdot3\,960\}\,2^{n-10}
 =\frac{495}{64}\,2^n<19\cdot2^n.\qedhere
\]
\end{proof}

\begin{table}[tb]
\centering
\caption{The certificates for the $\nicefrac{17}{20}$-bound with minimum period at least $4$. Maximum length is the length of a longest word in each set. Files are in \code{packing\_plans/proof\_17\_20/}, relative to the software repository (the file names are clickable). The word total counts distinct words across files. Check times include dominance-box recomputation and SMT verification.}
\label{tab:certificates-four}
\begin{tabular}{@{}rccrrlr@{}}
\toprule
Dimension & Quantity bounded & Domain & Words & Maximum length & File & Check time\\
\midrule
1 & Density & $\mathcal D_1^{(4)}$ & 1 & 1 & \href{https://github.com/grzegorz-gutowski/pinwheel_scheduling/blob/main/packing_plans/proof_17_20/D1.txt}{\code{D1.txt}} & $<1$\,s\\
2 & Density & $\mathcal D_2^{(4)}$ & 1 & 2 & \href{https://github.com/grzegorz-gutowski/pinwheel_scheduling/blob/main/packing_plans/proof_17_20/D2.txt}{\code{D2.txt}} & $<1$\,s\\
3 & Density & $\mathcal D_3^{(4)}$ & 1 & 4 & \href{https://github.com/grzegorz-gutowski/pinwheel_scheduling/blob/main/packing_plans/proof_17_20/D3.txt}{\code{D3.txt}} & $<1$\,s\\
4 & Density & $\mathcal D_4^{(4)}$ & 1 & 4 & \href{https://github.com/grzegorz-gutowski/pinwheel_scheduling/blob/main/packing_plans/proof_17_20/D4.txt}{\code{D4.txt}} & $<1$\,s\\
5 & Density & $\mathcal D_5^{(4)}$ & 26 & 186 & \href{https://github.com/grzegorz-gutowski/pinwheel_scheduling/blob/main/packing_plans/proof_17_20/D5.txt}{\code{D5.txt}} & $<1$\,s\\
6 & Density & $\mathcal D_6^{(4)}$ & 148 & 792 & \href{https://github.com/grzegorz-gutowski/pinwheel_scheduling/blob/main/packing_plans/proof_17_20/D6.txt}{\code{D6.txt}} & $<1$\,s\\
7 & Density & $\mathcal D_7^{(4)}$ & 719 & 2\,304 & \href{https://github.com/grzegorz-gutowski/pinwheel_scheduling/blob/main/packing_plans/proof_17_20/D7.txt}{\code{D7.txt}} & 6\,s\\
8 & Density & $\mathcal D_8^{(4)}$ & 2\,665 & 4\,752 & \href{https://github.com/grzegorz-gutowski/pinwheel_scheduling/blob/main/packing_plans/proof_17_20/D8.txt}{\code{D8.txt}} & 66\,s\\
9 & Density & $\mathcal D_9^{(4)}$ & 8\,284 & 3\,420 & \href{https://github.com/grzegorz-gutowski/pinwheel_scheduling/blob/main/packing_plans/proof_17_20/D9.txt}{\code{D9.txt}} & 242\,s\\
9 & Potential & $\mathcal P_9^{(4)}$ & 22\,752 & 3\,960 & \href{https://github.com/grzegorz-gutowski/pinwheel_scheduling/blob/main/packing_plans/proof_17_20/P9.txt}{\code{P9.txt}} & 821\,s\\
10 & Potential & $\mathcal P_{10}^{(4)}$ & 3\,747 & 3\,348 & \href{https://github.com/grzegorz-gutowski/pinwheel_scheduling/blob/main/packing_plans/proof_17_20/P10.txt}{\code{P10.txt}} & 107\,s\\
\midrule
\multicolumn{3}{@{}l}{Total} & 32\,136 & & & 1\,242\,s\\
\bottomrule
\end{tabular}
\end{table}

\subparagraph*{Use of AI.}
We used OpenAI Codex to assist with manuscript revision, reference checking, and mathematical exposition.
Generative AI tools were not used to develop the research programs.
Some ideas from the communicated AI-generated proofs informed our approach; these ideas were subsequently traced to earlier research, which is credited above.
The authors take full responsibility for the content of this paper.

\section{Discussion}
\label{sec:discussion}
A natural next question is whether every real-valued instance with minimum period at least $5$ and density at most $\nicefrac{13}{15}$ is schedulable.
This would establish the $a_1=5$ case of Kawamura's \hyperref[conj:real_a_1]{conjecture} and extend it to real periods.
Computations searching for suitable density and potential certificates are ongoing.

\bibliographystyle{plainurl}
\bibliography{pinwheel-scheduling}
\end{document}